\documentstyle{agile7th}

\input epsf.sty
\input epsfig.sty
\input psfig.sty

\def \AAP #1 #2 {{\em Astron. Astrophys.\/} {\bf #1}, #2}
\def \AAL #1 #2 {{\em Astron. Astrophys. Lett.\/} {\bf #1}, L#2}
\def \AAR #1 #2 {{\em Astron. Astrophys. Rev.\/} {\bf #1}, #2}
\def \AAS #1 #2 {{\em Astron. Astrophys. Suppl. Ser.\/} {\bf #1}, #2}
\def \AJ #1 #2 {{\em Astron. J.\/} {\bf #1}, #2}
\def \ANNREV #1 #2 {{\em Ann. Rev. Astron. Astrophys.\/} {\bf #1}, #2}
\def \APJ #1 #2 {{\em Astrophys. J.\/} {\bf #1}, #2}
\def \APJL #1 #2 {{\em Astrophys. J. Lett.\/} {\bf #1}, L#2}
\def \APJS #1 #2 {{\em Astrophys. J. Suppl.\/} {\bf #1}, #2}
\def \APSS #1 #2 {{\em Astrophys. Space Sci.\/} {\bf #1}, #2}
\def \ASR #1 #2 {{\em Adv. Space Res.\/} {\bf #1}, #2}
\def \BAIC #1 #2 {{\em Bull. Astron. Inst. Czechosl.\/} {\bf #1}, #2}
\def \JSQRT #1 #2 {{\em J. Quant. Spectrosc. Radiat. Transfer\/} {\bf #1}, #2}
\def \MN #1 #2 {{\em Mon. Not. R. Astr. Soc.\/} {\bf #1}, #2}
\def \MEM #1 #2 {{\em Mem. R. Astr. Soc.\/} {\bf #1}, #2}
\def \PLR #1 #2 {{\em Phys. Lett. Rev.\/} {\bf #1}, #2}
\def \PASJ #1 #2 {{\em Publ. Astron. Soc. Japan\/} {\bf #1}, #2}
\def \PASP #1 #2 {{\em Publ. Astr. Soc. Pacific\/} {\bf #1}, #2}
\def \NAT #1 #2 {{\em Nature\/} {\bf #1}, #2}
\def \SAIT #1 #2 {{\em Mem.\ Soc.\ Astron.\ It.\/} {\bf #1}, #2}
\def \MESS #1 #2 {{\em The Messenger\/} {\bf #1}, #2}
\def \ASTRNACH #1 #2 {{\em Astron. Nach.\/} {\bf #1}, #2}
\def \AGPSR #1 #2 {{\em ASI Special Publication\/} {\bf #1}, #2}

\begin{opening}

\title{Is the  Gamma Ray Bursts emission suppressed at high energy?}
\author{D.Guetta$^{1}$, E.Pian$^{2,3,4}$}
\institute{$^1$Osservatorio astronomico di Roma, v. Frascati 33,
00040 Monte Porzio Catone, Italy\\
$^2$ Osservatorio Astronomico di Trieste, Via G.B. Tiepolo, 11 - 34143 Trieste, Italy \\
$^3$ Scuola Normale Superiore, Piazza dei Cavalieri 7, I-56126 Pisa, Italy \\
$^4$ European Southern Observatory, Karl-Schwarzschild-Strasse 2
D-85748 Garching bei M\"unchen, Germany}
\date{} 
\end{opening}

\begin{document}

\oddpagefooter{}{}{} 
\evenpagefooter{}{}{} 
\medskip  

\begin{abstract} 
We compare the luminosity function and rate inferred from the GBM  long bursts peak flux distribution with those inferred from the Swift and BATSE peak flux distribution. We find that  the GBM, BATSE and the Swift peak fluxes can be fitted by  the same luminosity function implying the consistency of these three samples. Using the trigger algorithm of the LAT instrument we derive important  information on the flux at 100 MeV compared to lower energy detected by the GBM. We find that the simple extension  of the synchrotron emission to high energy cannot justify the low rate of GRBs detected by LAT and for several GRBs detected by the GBM, the flux at $>100$  MeV should be suppressed.  Two bursts, GRB090217 and GRB 090202b,
 detected by LAT have very soft spectra in the GBM and therefore their high energy emission cannot be due to an extension of the synchrotron.
\end{abstract}

\medskip

\section{Introduction}
Gamma ray bursts (GRBs) are  the most powerful events in the
universe, their  total emitted energy outputs exceeding sometime $10^{54}$ ergs, owing to relativistic aberration. In GRBs in fact,  the most extreme relativistic regimes are  attained among all high energy sources of macroscopic size.  Most of the energy is radiated in gamma-rays of 100 to 1000 keV,  with tails up to the GeV domain, as detected formerly by CGRO/EGRET (Dingus 1995, Hurley et al. 1994, Gonzalez et al. 1994) and more recently and with much better detail by the AGILE/GRID and Fermi/LAT instruments (Marisaldi et al. 2009, Giuliani et al. 2008,  Giuliani et al. 2009, Abdo et al. 2009, 
 Abdo et al. 2009, Bissaldi et al. 2009, Granot 2009).   
 
Long GRBs  ($T_{90}>2$s) are thought to trace the history of massive star formation of the Universe and are detected all the way from locally (40 Mpc)  to the edge of the Universe ($z \sim 8$, Salvaterra et al. 2009; Tanvir et al. 2009), so that they are ideal targets for the unbiased study of cosmological effects on the propagation of high energy photons and on the evolution of star formation. 

However the number of GRBs with a measured redshift is still
limited and at present we cannot derive directly the GRB luminosity
function and rate evolution that are fundamental to understand the
nature of these objects.
We can constrain the luminosity function and rate distribution by
fitting their   peak flux distributions  to those
expected for a given luminosity function and GRB rate, as done for CGRO/BATSE and {\it Swift}/BAT GRBs (Piran 1992,
Cohen \& Piran 1995, Fenimore \& Bloom 1995, Loredo \& Wasserman
1995, Horack \& Hakkila 1997, Loredo \& Wasserman 1998, 
Schmidt 1999, Schmidt 2001, Sethi \& Bhargavi 2001, Guetta, Piran \& Waxman
2005, Guetta \& Piran 2005, 2006, 2007). Since the observed flux
distribution is a convolution of these two unknown functions we must
assume one and find a best fit for the other. 

Here we concentrate on the $\sim$ 250 long GRBs detected so far by the Gamma Ray Burst Monitor 
(GBM) onboard the Fermi GST (Table 1).  We assume that the
rate of long bursts follows the star formation rate and we
search for the parameters of the luminosity function. 
In the first part of this paper, we show that one can obtain a fully consistent
fit for  the GBM, BATSE and the {\it Swift} peak flux populations, implying that 
the GBM sample has properties similar to those of BATSE and {\it Swift}.

Only 10 long bursts have been detected above  30 MeV by the LAT.  
Considering only the ratio between the LAT and GBM field of view (2.5 sr and more than 8 sr, respectively) and assuming comparable sensitivity of the 2 instruments in their energy ranges to the level of emission expected from GRBs, we would have expected
that $\sim 1/3$ of the GBM bursts has a $> 30$ MeV counterpart and not only $\sim$ 5\%, as detected.  There could be 2 effects at play: the LAT sensitivity and the fact that the intrinsic brightness of the very high energy tails of GRBs can be lower than predicted with a simple extrapolation of the soft gamma-ray spectrum or based on synchrotron and inverse Compton flux estimates (see e.g. Ando, Nakar \& Sari 2008, who have predicted a LAT detection rate of about 20 GRB/yr and see Fan 2009).  Another possible explanation not related to instrumental or intrinsic GRB physics reasons is that in very distant GRBs with highly collimated jets, the intrinsic strong high
energy emission is suppressed by the diffuse extragalactic infrared background (Gilmore,  Prada,  \& Primack  2009).

The statistics of the LAT detection with respect to GBM detection is not dissimilar from that of the GRO instruments EGRET vs BATSE:  if the detection were only related to the FOV extension, EGRET should have detected about 50 GRBs out of the nearly 3000 detected by BATSE (BATSE covered virtually all sky, while the EGRET FOV was 30 deg in diameter).  The fact that EGRET detected only about 20 GRBs above 20 MeV, of which about 5 with the spark chamber at the higher energies (i.e. higher than 200 MeV), reflects broadly the LAT-vs-GBM statistics and indicates an intrinsic paucity of detected very high energy tails.    Similar to the LAT-detected GRBs, the EGRET-detected ones were among the brightest BATSE GRBs (Dingus et al. 1995).  Recently,  Kaneko et al. (2008) analyzed the spectral shapes of the BATSE GRBs observed by the TASC calorimeter, i.e. with  MeV emission, and found that the spectra of these bursts are quite hard  (i.e. have fairly high E$_{\rm peak}$ or small high energy spectral index). This is similar to what we find for the bursts detected by LAT (see Fig. 1).

A proper understanding of the  link  between the GRB spectral maximum emission (100 keV  - 1 MeV) and its countepart above few MeV  is the key to get insight into the inner mechanism of power generation and into the jet formation and collimation (see Kumar \& Barniol Duran 2009,
 Zhang \& Pe'er 2009,  Zou, Fan, Piran 2009,  Li 2009), and ultimately will explain  how the GRB emission at the highest energies correlates with the fundamental GRB parameters (Amati, Frontera, Guidorzi 2009,   Ghirlanda, Nava, \& Ghisellini 2009).

In this paper, after having verified that the properties of the GBM GRB population do not differ from those of the previous GRB missions BATSE and Swift-BAT (Section 2),   we analyze the sensitivity properties of the LAT and we estimate (following Band et al. 2009) the LAT detection rate of the high energy (100 MeV - 10 GeV) counterparts of GBM GRBs, under the hypothesis of  a simple extrapolation of the GRB spectrum to energies larger than 1 MeV.

\begin{figure}[ht!]
\centerline{\psfig{figure=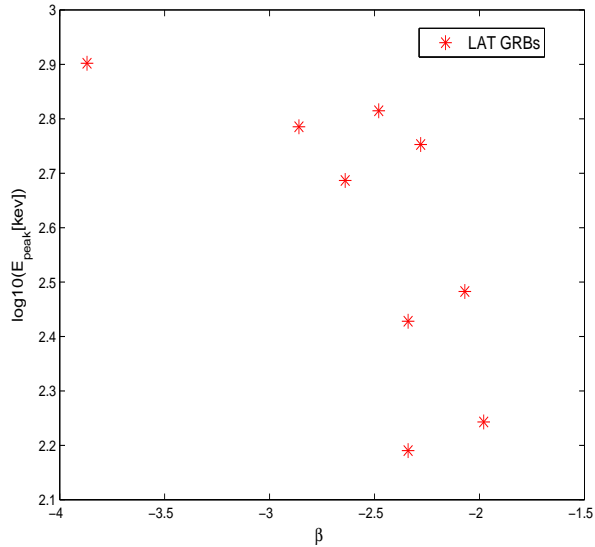,height=8cm,width=9.cm}}
\caption{The spectral parameters, $\beta$ and $E_{p}$ of the GRBs detected by LAT}
\end{figure}

\section{The luminosity function  from the GBM sample}
\subsection{The GBM, BATSE and {\it Swift}-BAT samples}

Our  methodology follows Guetta et al. (2005) and Guetta \& Piran (2006).
We consider all the long ($T90 > 2$ seconds) bursts detected by the GBM until September 2009 and compare their peak fluxes (when measured) with the peak fluxes of the long GRBs detected
by BATSE and Swift. The comparison is done in the energy band 50-300 keV, which is the
band where BATSE detects GRBs. 
Note that for this comparison we have to convert 
GBM peak fluxes reported in different bands to fluxes in the 50-300 keV
band. We have done this conversion by assuming  that  
the spectrum around the maximum is fitted by the same spectral model that has been adopted to fit the average spectrum with identical fit parameters.  

The average spectrum parameters of the GBM bursts 
are $<\alpha>\sim -0.8$ $<\beta>\sim -2.33$ and $E_p\sim 217 $ keV (only for GRBs fitted with the Band law). Using these average parameters, we estimate, from Band (2003), a GBM  sensitivity  in the 50-300 keV of  $P^{(50-300)keV}_{\rm lim, GBM}\sim$ 0.5 ph cm$^{-2}$ s$^{-1}$.   Hereafter, we  consider for our analysis only the  long ($T_{90} > 2$ s) GBM bursts with peak flux higher than that threshold, consisting in a sample of 125 bursts.

In  the BATSE sample we have included
all the long GRBs detected while the BATSE onboard trigger was set
to a significance of  5.5 $\sigma$ over background in at least two detectors, in the
energy range 50-300 keV. Among those we selected the bursts for which the peak
flux in a 1024 ms timescale is higher than the BATSE threshold 
for long bursts reported by Band (2003),
 $P^{(50-300)keV}_{\rm lim, BATSE}\sim$ 0.25 ph cm$^{-2}$ s$^{-1}$.
This yields a sample of 1425 bursts.

For {\it Swift} we consider all long bursts detected until September
2009  in the energy range 15-150 keV.
We convert the BAT 15-150 keV peak fluxes to fluxes in the 50-300 keV
band using the BAT peak fluxes and spectral
parameters provided by the Swift team\footnote{See {\it Swift}
information page
http://swift.gsfc.nasa.gov/docs/swift/archive/grb\_table.html}. 
We consider only the bursts with peak fluxes above the 
threshold of $P^{(50-300)keV}_{\rm lim, Swift}\sim
0.3$ ph cm$^{-2}$ s$^{-1}$ 
(Band 2003), which is equivalent to a sample of 259 GRBs.
The fluxes of the three instruments are shown in Fig. 2, where we have normalized to 
the number of GRBs in our GBM sample.

\begin{figure}[ht!]
\centerline{\psfig{figure=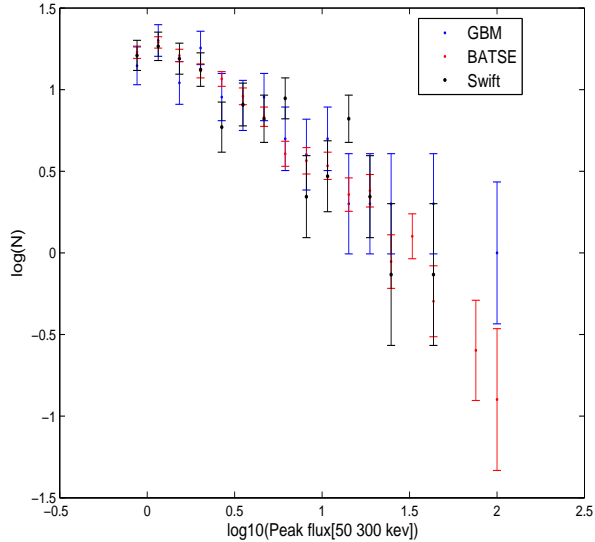,height=8cm,width=9.cm}}
 \caption{ The logN-logS distributions of the three instruments in the 50-300
kev band}
\end{figure}

\subsection{Luminosity function and comoving rate}

We now construct a luminosity function that we will normalize to our GBM GRB distribution.
The method used to derive the luminosity function is essentially
the one used by Schmidt (1999), Guetta et al. (2005) and Guetta and Piran (2007). 
We consider a broken power law between lower and  upper limits which are
factors of $1/\Delta_1$ and $\Delta_2$ respectively times the
break luminosity $L^*$. The luminosity function (of the  peak
luminosity $L$) in the interval $\log L$ to $\log L + d\log L$ is:
\begin{equation}
\label{Lfun}
\Phi_o(L)=c_o
\left\{ \begin{array}{ll}
(L/L^*)^{\alpha} &  L^*/\Delta_1 < L < L^* \\
(L/L^*)^{\beta} & L^* < L < \Delta_2 L^*
\end{array}
\right. \;,
\end{equation}
where $c_o$ is a normalization constant so that the integral over
the luminosity function equals unity. We stress that the
luminosity considered here is the ``isotropic" equivalent
luminosity, which is the one relevant for detection. It does not
include a correction factor due to beaming.

Assuming that long GRBs follow the star formation rate we employ
four parametrizations of the star formation
rate:\hfill\break (i) Model SF2 of Porciani \& Madau (2001):
\begin{equation}
\label{SFR} R_{GRB}(z)=R_{\rm SF2}(z)=  \rho_0 \frac{23
\exp(3.4z)}{\exp(3.4z)+22}
\end{equation}
where $\rho_0$ is the GRB rate at $z=0$.

An important factor  is the cosmological k correction. We
approximate the typical effective spectral index in the observed
range of 50 keV to 300 keV as ($N(E)\propto E^{-1.6}$). 
The use of this average correction is justified when we
compare estimates of the luminosity based on this average value and
on the real spectrum (see Guetta and Piran 2007).

\subsection{Monte Carlo simulations}

In our Monte Carlo simulations, each generated GRB is given a redshift, a luminosity
and a spectrum according to the specific intrinsic distributions that have been described above.
For each burst we compute the observed peak flux and compare the peak flux distribution
with the observed one.
We find the best-fitting LF parameters and their dispersion by
$\chi^2$ minimization. We  vary the luminosity function parameters
$\alpha$, $\beta$, and $L^*$ keeping $\Delta_1=100$ and
$\Delta_2=100$ and inspect the quality of the fit to the observed
GBM peak flux distribution. Once we obtain the best fit parameters
for the GBM sample we test the quality of the fit with both the
observed BATSE and the {\it Swift} peak flux distributions. We then
repeat the same procedure and look for the luminosity function
parameters that best fit the {\it Swift} sample and the BATSE sample. 
Then we check the
quality of the fit with these parameters with the observed GBM, BATSE and
the {\it Swift} peak flux distributions.  
The results of the fit are reported in Table 2. These show that the
best fit parameters $\alpha$ and $\beta$ are rather robust and the
values of $\alpha$ and $\beta$ found for these different samples
are all consistent within the error bars. 

We consider a LF that fit quite well all the samples (we call it LFb), this has $\alpha=1$,
$\beta=2.7$ and $L^*=5\times 10^{51} $ erg cm$^{-2}$ s$^{-1}$.
Fig.  3, 4 and 5 depict a
comparison of the peak flux distributions
  of GBM, BATSE and {\it Swift}  with the predicted distribution
obtained with the LF given above.

\begin{figure}[ht!]
\centerline{\psfig{figure=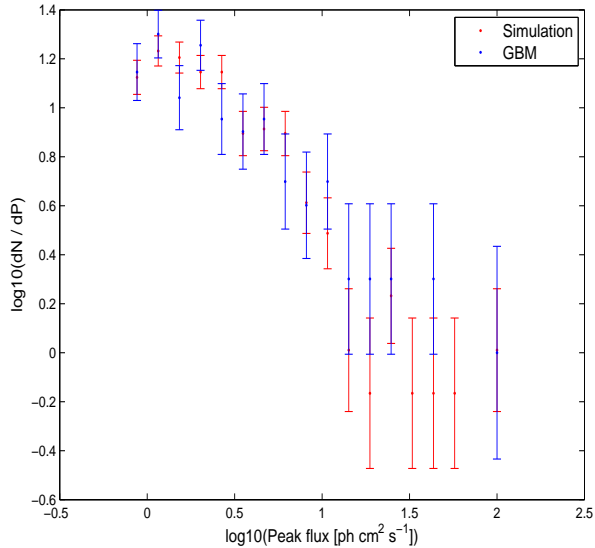,height=8cm,width=9.cm}}
\caption{The predicted peak flux
distribution  with the luminosity function
parametrs of LFb compared withe GBM fluxes}
\end{figure}

\begin{figure}[ht!]
\centerline{\psfig{figure=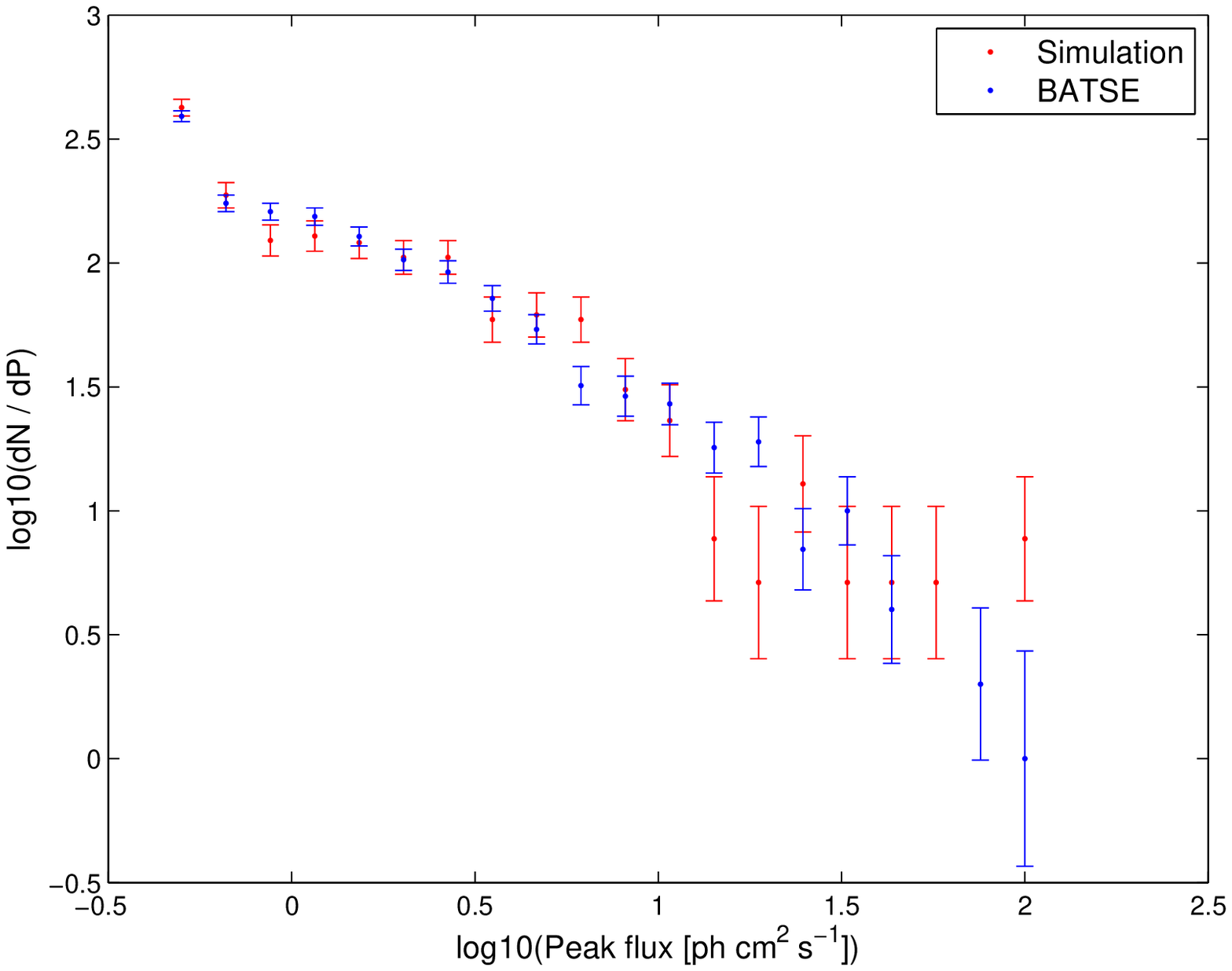,height=8cm,width=9.cm}}
\caption{The predicted peak flux
distribution  with the luminosity function
parametrs of LFb compared withe BATSE fluxes}
\end{figure}

\begin{figure}[ht!]
\centerline{\psfig{figure=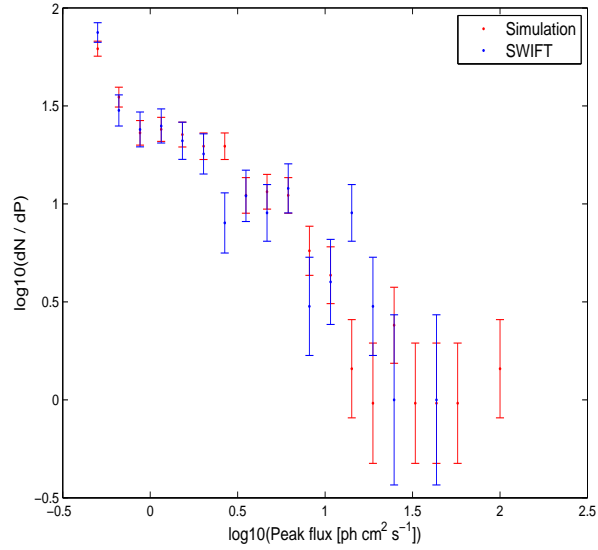,height=8cm,width=9.cm}}
\caption{The predicted peak flux
distribution  with the luminosity function
parametrs of LFb compared withe {\it Swift} fluxes}
\end{figure}

\section{Estimation of the LAT Detection Sensitivity to GRBs }

Following Band et al. 2009 we compute the LAT's burst detection sensitivity
using a semi-analytical approach based on the likelihood ratio test
introduced by Neyman \& Pearson (1928). This test is applied extensively to
photon-counting experiments (Cash 1979) and has been used to
analyze the gamma-ray data from COS-B  (Pollock 1981, 1985)
and EGRET, (Mattox 1996). The statistic for this test is the
likelihood for the null hypothesis for the data divided by the
likelihood for the alternative hypothesis, here that burst flux is
present. This methodology is the basis of the likelihood tool 
 used to analyze LAT observations; here we perform a
semi-analytic calculation for the simple case of a point source on a
uniform background.

The important quantity to write is the
\begin{eqnarray}
T_{S}  & = & 2~T_{obs}\int_{\Delta\Omega}\int_{E_1}^{E_2} A_{eff}(E) B(E)
   \left[ \left(1+G(E,\Omega)\right)  \right. \nonumber \\
   & &\left.  \ln\left(1+G(E,\Omega)\right) -G(E,\Omega) \right] dE d\Omega
\label{eq:Ts}
\end{eqnarray}
where we have defined a signal-to-noise ratio
$G(E,\Omega)=S(E)F(E,\Omega)/B(E)$,  $E_1=100$ MeV
and $E_2=10$ GeV.
Here we assume $S(E)\sim N_0 E^{-\delta}$  between 100 MeV and 10 GeV,
where $N_0$ is determined from the flux at 100 MeV.
$A_{eff}$ is the effective area given in Atwood et al. 2009. This area decreases very rapidly with  increasing  boresight angle and we take this into account.

We have estimated the flux at 1 MeV for all our GRBs that have a measured fluence\footnote{Some GRBs with no reported peakflux have a  fluence measurement.  While these GRBs are excluded from the peak flux analysis described above, they have been included in our estimate of the MeV flux.}, a measured boresight angle and reliable spectral fit parameters.   For these bursts we extrapolate  the flux at 1 MeV using the spectrum, normalized with the fluence. 
For GRBs with multi-peak structure, for which the spectral fit parameters and relevant observed quantities are reported individually for each pulse, we have estimated the 1 MeV flux using the spectrum of the last pulse (GRBs 081009, 090509, 090516A, and 090610B).

We applied one caveat for GRBs with spectra fitted in the GBM band with single power-laws: in this case, the spectrum cannot be extrapolated straightforwardly to 1 MeV  if the energy range used for the fit is much softer.  Thus, for the 1-MeV flux estimate we have retained the GRBs fitted with single power-laws over a range whose upper energy is larger than 800 keV, i.e. close enough to 1 MeV to guarantee that the extrapolation will not dramatically overestimate the flux at 1 MeV if there is a spectral break beyond the spectral fitting range.  Following this criterion, we excluded GRBs 080818A, 080928, 081206C, 081225, for which the 1-MeV flux has been automatically set to zero.   In total, we could reliably estimate the flux at 1 MeV for 167  GRBs.

 Analogously, we have estimated by extrapolation the flux at 100 MeV.  However, the GRBs for which this extrapolation results in a non-zero flux are only  those fitted with the Band law.  In fact,  when the  flux at 100 MeV is extrapolated from spectra fitted with  power-law plus exponential high energy cutoff, the result is generally null, because of the high suppression caused by the exponential at MeV energies.  On the other hand, the extrapolation of spectra fitted by single power-law in the GBM range to 100 MeV is very likely to overestimate the flux at 100 MeV by many orders of magnitude. The 100 MeV fluxes of GRBs fitted with single power-laws have been arbitrarily set to zero.  Twelve  (number may change) GRBs fitted with the Band law have $\beta > -2$, which translates in a $\nu F_{\nu}$ spectrum that rises with energy.  The spectral fit for these GRBs extends to 1 MeV, so that the flux estimate at this energy is reliable.  However,  it is likely that the spectrum curves at higher energies, starting well below 100 MeV, so that the flux estimate at 100 MeV with the $\beta$ index of the formal Band-law fit may overestimate the real flux by orders of mangitude.  For these GRBs, we have estimated the 100-MeV flux under the assumption that the spectrum above 1 MeV steepens\footnote{Note that a spectral steepening  at an energy comprised between 1 and 100 MeV may occur in all our GRBs, and not only in those whose GBM high energy spectrum is fitted with $\beta > -2$.  However,  for GRBs with $\beta \le -2$, ignoring this possible steepening leads to a less dramatic discrepancy.} to 
$\beta = -2.3$ (the average value of the whole sample of Band-fitted GRBs).  In the tests we will implement in the following, we will both include and exclude those 12 GRBs and will provide the results in both cases.       

The Point Spread Function (PSF), $F(E,\Omega)$, is taken from Burnett 2007 (private comunication):

\begin{equation}
f(\gamma,u)=(1-\frac{1}{\gamma})(1+\frac{u}{\gamma})^{-\gamma}
\end{equation}

with $\Omega=2\pi\sigma^2 u$ where 
$\sigma(E)=\sqrt{(a^2+(b*(E/100)^{-0.8})^2} $ (in radiant)
with a=58e-3 e b=377e-6 (on-ground analysis).
The parameter $\gamma$ depends very weakly on the energy and vary between 2-2.5.
We can consider $\gamma\sim 2$ for our analysis (Burnett 2007).

Like Band 2009, we assume a
spatially uniform background with a power law spectrum
\begin{equation}
B(E)=B_{0}\left(\frac{E}{\hbox{100 MeV}}\right)^{\gamma}
   \quad \hbox{ph
cm$^{-2}$ MeV$^{-1}$ s$^{-1}$ sr$^{-1}$}
 \label{be}
\end{equation}
where the value of the normalization constant $B_{0}$ is
set to mimic the expected background rate. For modeling the
on-ground trigger the background rate above 100 MeV is set to 4~Hz. 
The spectral index is set
to be $\gamma=-2.1$. 
In our modeling we assume that the GRB flux is constant over a
duration $T_{GRB}$.  Since we seek the optimal detection
sensitivity, we calculate $T_S$ for $T_{obs}=T_{GRB}$. 
We also consider the effect of the boresight angle, i.e. the effective area 
drops of a factor of $\sim 10$ between $10^0$ and $70^0$ of boresight angle.

The expected number of counts from a burst flux S(E) is

\begin{equation}
N_S = T_{\rm obs}\int_{\Delta\Omega}\int_{E_1}^{E_2} A_{eff}(E)S(E)F(E,\Omega)
\label{ns}
\end{equation}

We require
$T_{S}\ge25$ and at least 10 source counts in the LAT
detector, corresponding to a threshold significance of
5$\sigma$ and a minimum number of GRB counts.

In the sample there are 84 bursts that can be fitted with a Band spectra and have a determined boresight angle, 34 GRBs have 
$\theta>80^0$ and therefore cannot be detected by LAT. We are left with 50 bursts and 28 of them were not detected by LAT.
 For these 28 bursts the flux S(E) in the eq.\ref{eq:Ts}  and \ref{ns} between 100 MeV-and 10 GeV
is given by  $S(E)\sim N_0 E^{-\beta}$  where $N_0$ is determined from the flux at 100 MeV obtained by extrapolating  the  Band spectra used to fit the GBM data at high energy, using the spectral index $\beta$ from the data (or -2.3 if $\beta>-2$) .
An interesting quantity that can be defined for these bursts is the ratio, s,  between the flux at 100 MeV obtained by simply extending the synchrotron emission at high energies, and the flux at 100 MeV that one needs in order to be at the threshold of the LAT detection.  
In Fig. 6 we plot this ratio as a function of the spectral index $\beta$ and in Fig.7 we plot this ratio
as a function of the peak energy. As it is clear from these figures, the softer  the spectrum, the smaller the value of s, implying that the LAT does not detect any signal,   as expected. It is interesting to note that there are $\sim 22$ bursts that could have been detected 
by LAT (have $s\geq 1$) but were not detected. This implies that there is a suppression in the flux at high energy and the spectrum is not simply the extension of the Band law.
In the same figures we also plot s vs $\beta$ ($E_{\rm peak}$) for the bursts detected by LAT. We see that for two bursts
GRB 090217 and GRB 090902B  that have very soft spectra, the value of $s$ is $<1$ and therefore the $>100 MeV$ emission cannot be an extrapolation of the synchrotron emission but another component
is needed to justify the emission at high energy.
Recently Bissaldi et al. (astro-ph/0909.2470) have shown that in the GRB 090202b a distinct extra
component is needed to explain the observed results in agreement with what we find. 
In the figures we also report the ratio between the measured flux at $\sim$ 100 MeV and the
flux extrapolated from the synchrotron emission for 080916c (the diamond mark).

\section{Discussion}

We have shown that the simple extension of the synchrotron emission cannot justify the lack of detection of high energy emission from GRBs. The emission is suppressed at high energy by some mechanism
(Fan 2009): i.e. the electrons are not accelerated to such an high energy, the high energy photons pair produce with low energy photons in the source and cannot escape the source or pair produce with external photons.

On the other hand  for the bursts detected by LAT there are two GRBs that have very soft synchrotron emission and their high energy emission cannot be explained by simple extrapolation of the low energy emission. There is evidence of another component at least in one GRB 090902b.

\begin{figure}[ht!]
\centerline{\psfig{figure=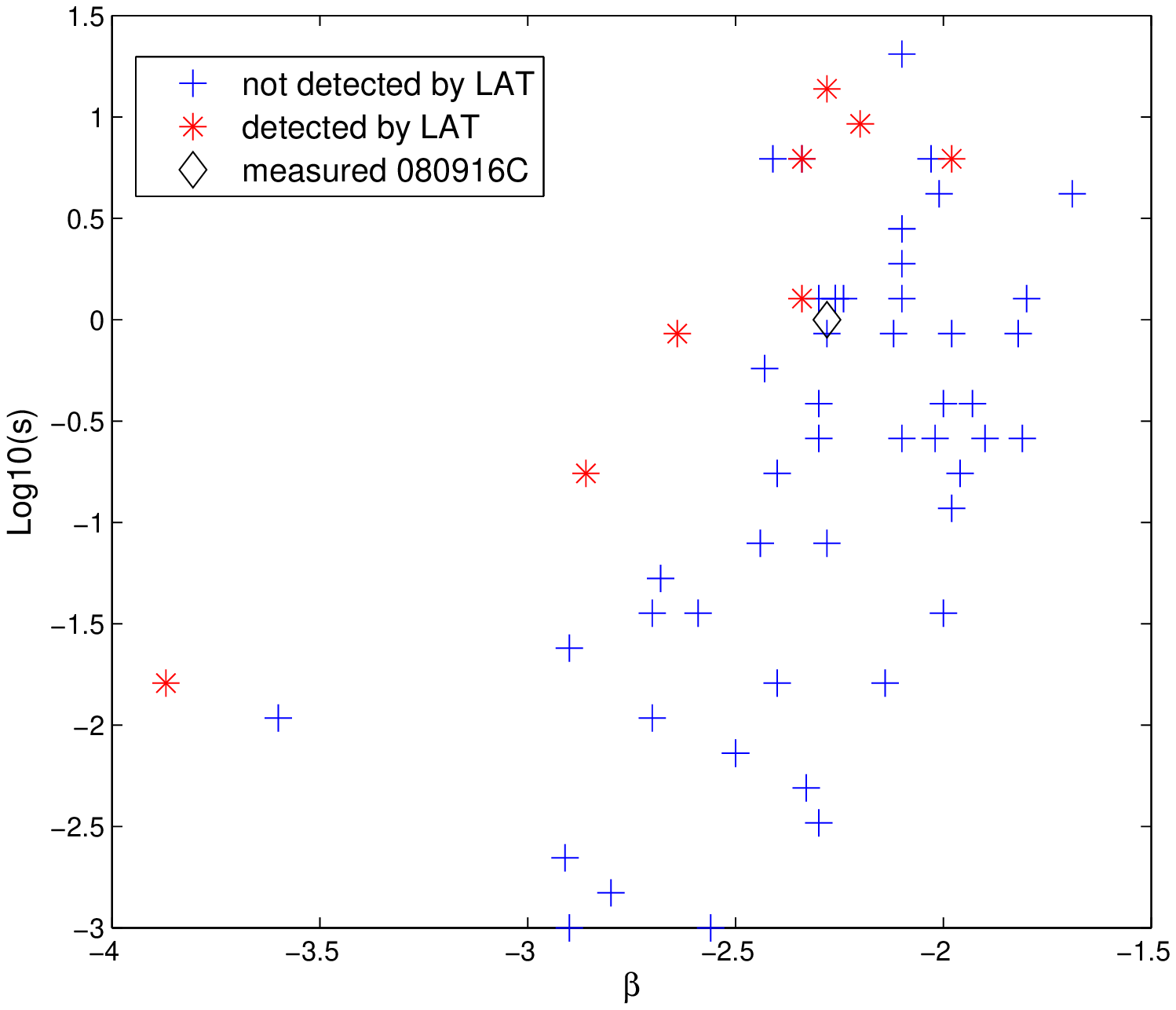,height=8cm,width=9.cm}}
 \caption{ The ratio s defined in the text as a function of the high energy spectral index $\beta$  for each burst not detected by LAT (blue cross), and detected by LAT (red strars). The measured value of 080916C is also reported (black diamond)}
\end{figure}

\begin{figure}[ht!]
\centerline{\psfig{figure=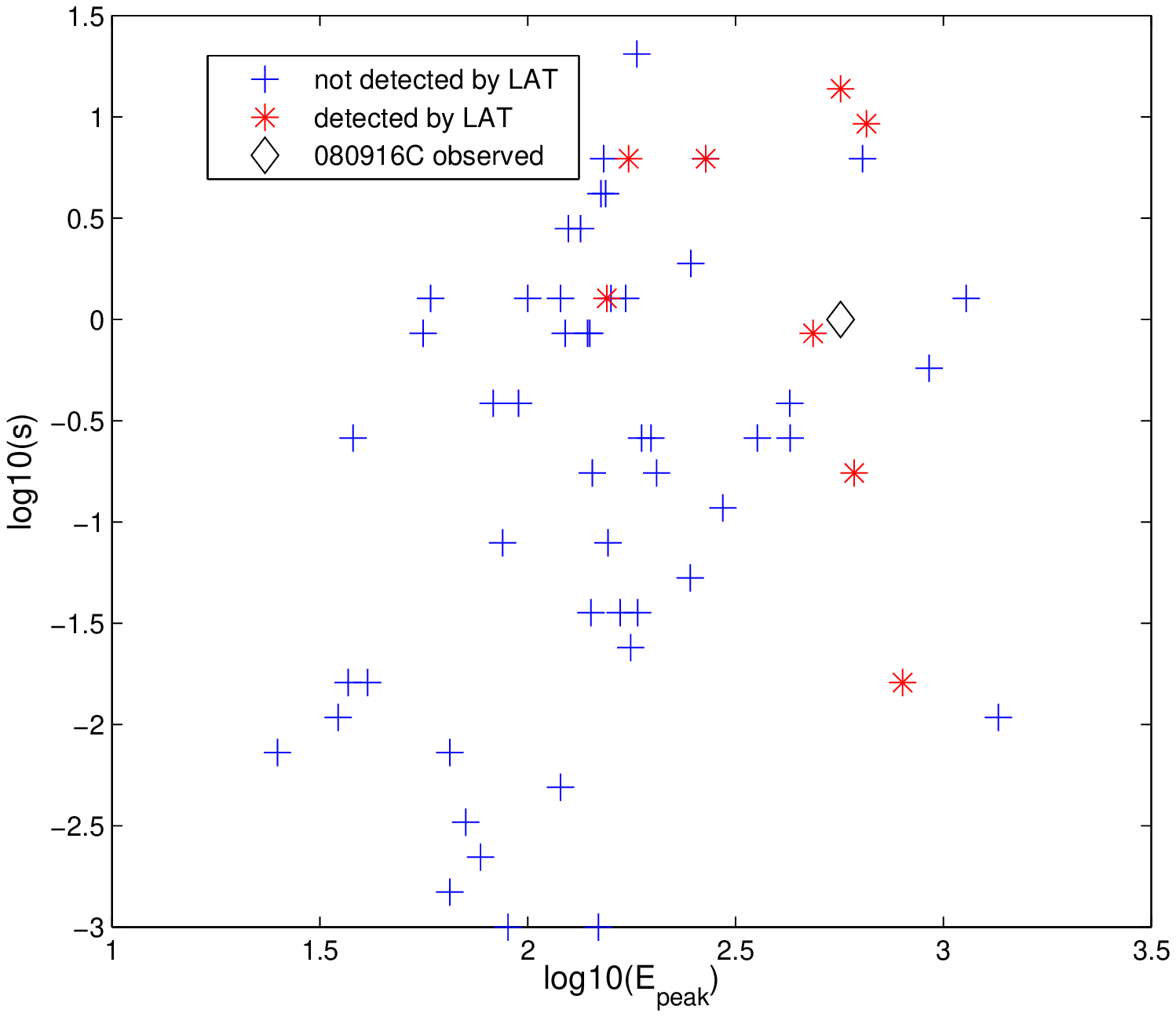,height=8cm,width=9.cm}}
 \caption{ The ratio s defined in the text as a function of the peak energy  for each burst not detected by LAT (blue cross), and detected by LAT (red strars). The measured value of 080916C is also reported (black diamond)}
\end{figure}


\onecolumn

\begin{table}[t]
\caption{GBM parameters for the bursts detected by Fermi. The first column report the GRB name, the second column the duration of the GRB, the third one the GRB fluence (set to -1 when not available), the fourth the peak flux (taken in 1 sec for long bursts and set to -1 when not available), the fifth column report the method used to fit the spectra, the sixth the peak energy (set to -1 when not available), the seventh and eight the $\alpha$ and $\beta$ spectral indexes (set to 1000 when not available). The ninth column report the angle, $\theta$,  from the LAT boresight, in deg (set to -1 when not available). The last column indicates LAT detection (1 = YES, 0 = NO)}
\begin{tabular}{lllllllllll}
 \hline
GRB & T90 &  Fluence (erg/cm2) & PF(ph/s/cm2) & Function & $E_{peak}$ & $\alpha$ & $\beta$ & $\theta$ & LAT \\ 
  &    sec &      $10^{-6}$erg/cm$^2$& ph/cm$^2$/sec& & keV & & & \\ 
  \hline
   080810  & 122.0  & 6.90e+00  & 1.85e+00  & PL+HEC  & 313.5  & -0.91  & -1000.00  & 61  & 0 \\ 
  080812  &  15.0  & -1.00e+00  & -1.00e+00  & PL+HEC  & 140.0  &  0.17  & -1000.00  & 71  & 0 \\ 
 080816A  &  70.0  & 1.86e+01  & 3.48e+00  & PL+HEC  & 146.7  & -0.57  & -1000.00  & 55  & 0 \\ 
 080816B  &   5.0  & -1.00e+00  & 1.38e+00  & PL+HEC  & 1230.0  & -0.37  & -1000.00  & 70  & 0 \\ 
 080817A  &  70.0  & -1.00e+00  & -1.00e+00  &      *  &  -1.0  & 1000.00  & -1000.00  & 80  & 0 \\ 
 080817B  &   6.0  & 2.60e+00  & -1.00e+00  &    SPL  &  -1.0  & -1.07  & -1000.00  & 68  & 0 \\ 
 080818A  &  50.0  & 2.26e+00  & -1.00e+00  &    SPL  &  -1.0  & -1.57  & -1000.00  & 68  & 0 \\ 
 080818B  &  10.0  & 1.00e+00  & -1.00e+00  & PL+HEC  &  80.0  & -1.30  & -1000.00  & 68  & 0 \\ 
  080823  &  46.0  & 4.10e+00  & -1.00e+00  & PL+HEC  & 164.7  & -1.20  & -1000.00  & 77  & 0 \\ 
  080824  &  28.0  & 2.30e+00  & -1.00e+00  &   Band  & 100.0  & -0.40  & -2.10  & 17  & 0 \\ 
 080825C  &  22.0  & 2.40e+01  & -1.00e+00  &   Band  & 155.0  & -0.39  & -2.34  & 60  & 1 \\ 
  080830  &  45.0  & 4.60e+00  & -1.00e+00  &   Band  & 154.0  & -0.59  & -1.69  & 23  & 0 \\ 
  080904  &  22.0  & 2.25e+00  & 3.50e+00  &   Band  &  35.0  &  0.00  & -2.70  & 23  & 0 \\ 
 080905A  &   1.0  & 2.80e-01  & 6.10e+00  &    SPL  &  -1.0  & -0.96  & -1000.00  & 28  & 0 \\ 
 080905B  & 159.0  & 4.10e-02  & 2.10e-01  &    SPL  &  -1.0  & -1.75  & -1000.00  & 82  & 0 \\ 
 080905C  &  28.0  & 4.60e+00  & 4.40e+00  & PL+HEC  &  78.8  & -0.90  & -1000.00  & 108  & 0 \\ 
 080906B  &   5.0  & 1.09e+01  & 2.20e+01  &   Band  & 125.3  & -0.07  & -2.10  & 32  & 0 \\ 
  080912  &  17.0  & 3.30e+00  & 4.10e+00  &    SPL  &  -1.0  & -1.74  & -1000.00  & 56  & 0 \\ 
 080913B  & 140.0  & 2.20e+00  & -1.00e+00  & PL+HEC  & 114.0  & -0.69  & -1000.00  & 71  & 0 \\ 
 080916A  &  60.0  & 1.50e+01  & 4.50e+00  & PL+HEC  & 109.0  & -0.90  & -1000.00  & 76  & 0 \\ 
 080916C  & 100.9  & 2.40e+02  & 6.87e+00  &   Band  & 566.0  & -0.92  & -2.28  & 48  & 1 \\ 
  080920  &  85.0  & 2.40e+00  & 1.29e+00  &    SPL  &  -1.0  & -1.42  & -1000.00  & 16  & 0 \\ 
  080925  &  29.0  & 9.70e+00  & -1.00e+00  &   Band  & 120.0  & -0.53  & -2.26  & 38  & 0 \\ 
  080927  &  25.0  & 5.70e+00  & 2.00e+00  &    SPL  &  -1.0  & -1.50  & -1000.00  & 75  & 0 \\ 
  080928  &  87.0  & 1.50e+00  & -1.00e+00  &    SPL  &  -1.0  & -1.80  & -1000.00  & -1  & 0 \\ 
 081003C  &  67.0  & 5.40e+00  & -1.00e+00  &    SPL  &  -1.0  & -1.41  & -1000.00  & 48  & 0 \\ 
 081006A  &   7.0  & 7.10e-01  & -1.00e+00  &   Band  & 1135.0  & -0.77  & -1.80  & 16  & 0 \\ 
 081006B  &   9.0  & 7.30e-01  & -1.00e+00  &    SPL  &  -1.0  & -1.30  & -1000.00  & 3  & 0 \\ 
 081007A  &  12.0  & 1.20e+00  & 2.20e+00  &    SPL  &  -1.0  & -2.10  & -1000.00  & 116  & 0 \\ 
 081007B  &   0.5  & -1.00e+00  & -1.00e+00  &      *  &  -1.0  & 1000.00  & -1000.00  & -1  & 0 \\ 
  081009  &   9.0  & 3.50e+01  & -1.00e+00  &   Band  &  47.4  &  0.10  & -3.42  & 67  & 0 \\ 
  081012  &  30.0  & 3.80e+00  & 2.00e+00  & PL+HEC  & 360.0  & -0.31  & -1000.00  & 61  & 0 \\ 
  081021  &  25.0  & 5.30e+00  & 4.20e+00  &   Band  & 117.0  &  0.11  & -2.80  & 125  & 0 \\ 
 081024B  &   0.8  & 3.40e-01  & 4.20e+00  & PL+HEC  & 1583.0  & -0.70  & -1000.00  & -1  & 1 \\ 
 081024C  &  65.0  & 4.00e+00  & 1.00e+00  &   Band  &  65.0  & -0.60  & -2.50  & 78  & 0 \\ 
  081025  &  45.0  & 7.10e+00  & 4.50e+00  &   Band  & 200.0  &  0.15  & -2.05  & -1  & 0 \\ 
 081028B  &  20.0  & 2.00e+00  & 6.90e+00  & PL+HEC  &  70.0  & -0.55  & -1000.00  & 107  & 0 \\ 
 081101A  &   0.2  & 1.60e-01  & -1.00e+00  &    SPL  &  -1.0  & -1.14  & -1000.00  & -1  & 0 \\ 
 081101B  &   8.0  & 1.60e+01  & 1.03e+01  & PL+HEC  & 550.0  & -0.62  & -1000.00  & 116  & 0 \\  \hline
 \end{tabular}
 \end{table}
 \pagebreak
 \newpage
\begin{table}[t]
\begin{tabular}{llllllllll}
 \hline
GRB name &   t90 &     fluence &       peakflux &    Function & E$_{\rm peak}$ &  $\alpha$ &    $\beta$ &       $\theta$ &  LAT\\
                     &    sec &      $10^{-6}$erg/cm$^2$& ph/cm$^2$/sec& keV & & & \\ 
\hline
 081102A  &  88.0  & 2.10e+00  & -1.00e+00  &   Band  &  72.0  &  0.44  & -2.36  & -1  & 0 \\ 
 081102B  &   2.2  & 1.12e+00  & 3.68e+00  &    SPL  &  -1.0  & -1.07  & -1000.00  & 53  & 0 \\ 
 081105B  &   0.2  & 2.28e-01  & 2.00e+01  &    SPL  &  -1.0  & -1.17  & -1000.00  & 87  & 0 \\ 
  081107  &   2.2  & 1.64e+00  & 1.10e+01  &   Band  &  65.0  &  0.25  & -2.80  & 52  & 0 \\ 
 081109A  &  45.0  & 6.53e+00  & 3.20e+00  & PL+HEC  & 240.0  & -1.28  & -1000.00  & -1  & 0 \\ 
  081110  &  20.0  & -1.00e+00  & -1.00e+00  &      *  &  -1.0  & 1000.00  & -1000.00  & 67  & 0 \\ 
  081113  &   0.5  & 1.07e+00  & 2.00e+01  &    SPL  &  -1.0  & -1.28  & -1000.00  & 60  & 0 \\ 
 081118B  &  20.0  & 1.12e-01  & 6.70e-01  &   Band  &  41.2  &  0.80  & -2.14  & 41  & 0 \\ 
  081119  &   0.8  & 4.10e-01  & 7.20e+00  &    SPL  &  -1.0  &  1.30  & -1000.00  & 86  & 0 \\ 
  081120  &  12.0  & 2.70e+00  & 5.10e+00  &   Band  &  44.0  &  0.40  & -2.18  & 84  & 0 \\ 
  081121  &  -1.0  & -1.00e+00  & -1.00e+00  &      *  &  -1.0  & 1000.00  & -1000.00  & -1  & 0 \\ 
 081122A  &  26.0  & 9.60e+00  & 3.00e+01  &   Band  & 158.6  & -0.63  & -2.24  & 21  & 0 \\ 
 081122B  &   0.3  & 7.90e-02  & 1.60e+00  &    SPL  &  -1.0  & -1.50  & -1000.00  & 52  & 0 \\ 
  081124  &  35.0  & 9.50e-02  & 6.70e-01  &   Band  &  22.8  & -0.60  & -2.83  & 86  & 0 \\ 
  081125  &  15.0  & 4.91e+01  & 2.70e+01  &   Band  & 221.0  &  0.14  & -2.34  & 126  & 0 \\ 
  081129  &  59.0  & 2.00e+01  & 1.40e+01  &   Band  & 150.0  & -0.50  & -1.84  & 118  & 0 \\ 
 081130A  &  -1.0  & -1.00e+00  & -1.00e+00  &      *  &  -1.0  & 1000.00  & -1000.00  & 90  & 0 \\ 
 081130B  &  12.0  & 1.30e+00  & 1.80e+00  & PL+HEC  & 152.0  & -0.77  & -1000.00  & 66  & 0 \\ 
 081204B  &   0.3  & 4.88e-01  & 1.63e+01  &    SPL  &  -1.0  & -1.18  & -1000.00  & 46  & 0 \\ 
 081204C  &   4.7  & 1.48e+00  & 7.20e+00  &    SPL  &  -1.0  & -1.40  & -1000.00  & 56  & 0 \\ 
 081206A  &  24.0  & 4.00e+00  & 2.40e+00  &   Band  & 151.0  &  0.13  & -2.20  & 102  & 0 \\ 
 081206B  &  10.0  & -1.00e+00  & -1.00e+00  &      *  &  -1.0  & 1000.00  & -1000.00  & 82  & 0 \\ 
 081206C  &  20.0  & 1.19e+00  & 7.60e-01  &    SPL  &  -1.0  & -1.35  & -1000.00  & 71  & 0 \\ 
  081207  & 153.0  & 1.06e+02  & -1.00e+00  &   Band  & 639.0  & -0.65  & -2.41  & 56  & 0 \\ 
  081209  &   0.4  & 5.90e-01  & 7.80e+00  &   Band  & 808.0  & -0.50  & -2.00  & 109  & 0 \\ 
  081213  &   0.0  & -1.00e+00  & -1.00e+00  &      *  &  -1.0  & 1000.00  & -1000.00  & 51  & 0 \\ 
 081215A  &   7.7  & 5.44e+01  & 6.89e+01  &   Band  & 304.0  & -0.58  & -2.07  & 86  & 1 \\ 
 081215B  &  90.0  & 2.80e+00  & -1.00e+00  & PL+HEC  & 139.0  & -0.14  & -1000.00  & 112  & 0 \\ 
  081216  &   1.0  & 3.60e+00  & 5.50e+01  &   Band  & 1235.0  & -0.70  & -2.17  & 95  & 0 \\ 
  081217  &  39.0  & 1.00e+01  & 4.00e+00  &   Band  & 167.0  & -0.61  & -2.70  & 54  & 0 \\ 
  081221  &  40.0  & 3.70e+01  & 3.30e+01  &   Band  &  77.0  & -0.42  & -2.91  & 78  & 0 \\ 
  081222  &  30.0  & 1.35e+01  & 1.48e+01  &   Band  & 134.0  & -0.55  & -2.10  & 50  & 0 \\ 
  081223  &   0.9  & 1.20e+00  & 2.20e+01  & PL+HEC  & 280.0  & -0.63  & -1000.00  & 28  & 0 \\ 
  081224  &  50.0  & -1.00e+00  & -1.00e+00  &      *  &  -1.0  & 1000.00  & -1000.00  & 16  & 0 \\ 
  081225  &  42.0  & 2.45e+00  & 6.00e-01  &    SPL  &  -1.0  & -1.51  & -1000.00  & 55  & 0 \\ 
 081226A  &   1.7  & 2.10e-01  & 3.20e+00  &    SPL  &  -1.0  &  1.17  & -1000.00  & 110  & 0 \\ 
 081226B  &   0.4  & 6.10e-01  & 1.77e+01  &   Band  & 300.0  & -0.20  & -1.82  & 22  & 0 \\ 
 081226C  &  60.0  & 2.32e+00  & 4.50e+00  & PL+HEC  &  82.0  & -1.04  & -1000.00  & 54  & 0 \\ 
  081229  &   0.5  & 8.70e-01  & 1.07e+01  &   Band  & 585.0  & -0.27  & -2.00  & 44  & 0 \\ 
\hline
 \end{tabular}
 \end{table}
 \pagebreak
 \newpage
\begin{table}[t]
\begin{tabular}{llllllllll}
 \hline
GRB name &   t90 &     fluence &       peakflux &    Function & E$_{\rm peak}$ &  $\alpha$ &    $\beta$ &       $\theta$ &  LAT\\
                     &    sec &      $10^{-6}$erg/cm$^2$& ph/cm$^2$/sec& keV & & & \\ 
\hline
081231  &  29.0  & 1.20e+01  & 1.53e+00  &   Band  & 152.3  & -0.80  & -2.03  & 20  & 0 \\ 
 090107B  &  24.1  & 1.75e+00  & 3.68e+00  & PL+HEC  & 106.1  & -0.68  & -1000.00  & -1  & 0 \\ 
 090108A  &   0.9  & 1.28e+00  & 3.97e+01  &   Band  & 104.8  & -0.47  & -1.97  & 60  & 0 \\ 
 090108B  &   0.8  & 7.90e-01  & 1.90e+01  &    SPL  &  -1.0  & -0.99  & -1000.00  & 72  & 0 \\ 
  090109  &   5.0  & 1.21e+00  & 2.76e+00  &    SPL  &  -1.0  & -1.50  & -1000.00  & 62  & 0 \\ 
 090112A  &  65.0  & 5.20e+00  & 7.00e+00  &   Band  & 150.0  & -0.94  & -2.01  & 4  & 0 \\ 
 090112B  &  12.0  & 5.40e+00  & 1.40e+01  &   Band  & 139.0  & -0.75  & -2.43  & 95  & 0 \\ 
 090117A  &  21.0  & 1.80e+00  & 9.60e+00  &   Band  &  25.0  & -0.40  & -2.50  & 51  & 0 \\ 
 090117B  &  27.0  & 2.10e+00  & 4.60e+00  &    SPL  &  -1.0  & -1.55  & -1000.00  & 49  & 0 \\ 
 090117C  &  86.0  & 1.10e+01  & 4.20e+00  &   Band  & 247.0  & -1.00  & -2.10  & 54  & 0 \\ 
 090126B  &  10.8  & 1.25e+00  & 4.90e+00  & PL+HEC  &  47.5  & -0.99  & -1000.00  & 18  & 0 \\ 
 090126C  &  -1.0  & -1.00e+00  & -1.00e+00  &      *  &  -1.0  & 1000.00  & -1000.00  & 68  & 0 \\ 
  090129  &  17.2  & 5.60e+00  & 8.00e+00  &   Band  & 123.2  & -1.39  & -1.98  & 22  & 0 \\ 
  090131  &  36.4  & 2.23e+01  & 4.79e+01  &   Band  &  58.4  & -1.27  & -2.26  & 40  & 0 \\ 
  090202  &  66.0  & 8.65e+00  & 7.77e+00  & PL+HEC  & 570.0  & -1.31  & -1000.00  & 55  & 0 \\ 
  090206  &   0.8  & 1.04e+00  & 1.90e+01  & PL+HEC  & 710.0  & -0.65  & -1000.00  & 72  & 0 \\ 
  090207  &  10.0  & 4.01e+00  & 1.88e+00  &    SPL  &  -1.0  & -1.59  & -1000.00  & 45  & 0 \\ 
  090213  &  -1.0  & -1.00e+00  & -1.00e+00  &      *  &  -1.0  & 1000.00  & -1000.00  & 17  & 0 \\ 
  090217  &  32.8  & 3.08e+01  & 1.12e+01  &   Band  & 610.0  & -0.85  & -2.86  & 34  & 1 \\ 
  090219  &   0.5  & 8.00e-01  & 7.70e+00  &    SPL  &  -1.0  & -1.43  & -1000.00  & 137  & 0 \\ 
  090222  &  18.0  & 2.19e+00  & 1.10e+00  &   Band  & 147.9  & -0.97  & -2.56  & 80  & 0 \\ 
 090227A  &  50.0  & 9.00e+00  & 4.57e+00  &   Band  & 1357.0  & -0.92  & -3.60  & 21  & 0 \\ 
 090227B  &   0.9  & 8.70e+00  & 3.46e+01  &   Band  & 2255.0  & -0.53  & -3.04  & 72  & 0 \\ 
 090228A  &   0.8  & 6.10e+00  & 1.33e+02  &   Band  & 849.0  & -0.35  & -2.98  & 16  & 0 \\ 
 090228B  &   7.2  & 9.96e-01  & 2.53e+00  & PL+HEC  & 147.8  & -0.70  & -1000.00  & 20  & 0 \\ 
 090301B  &  28.0  & 2.69e+00  & 4.40e+00  &   Band  & 427.0  & -0.98  & -1.93  & 56  & 0 \\ 
  090303  &  -1.0  & -1.00e+00  & -1.00e+00  &      *  &  -1.0  & 1000.00  & -1000.00  & -1  & 0 \\ 
  090304  &  -1.0  & -1.00e+00  & -1.00e+00  &      *  &  -1.0  & 1000.00  & -1000.00  & 31  & 0 \\ 
 090305B  &   2.0  & 2.70e+00  & 1.10e+01  &   Band  & 770.0  & -0.50  & -1.90  & 40  & 0 \\ 
 090306C  &  38.8  & 9.00e-01  & 2.40e+00  &   Band  &  87.0  & -0.32  & -2.28  & 14  & 0 \\ 
 090307B  &  30.0  & 1.70e+00  & 1.80e+00  & PL+HEC  & 212.0  & -0.70  & -1000.00  & 83  & 0 \\ 
 090308B  &   2.1  & 3.46e+00  & 1.42e+01  & PL+HEC  & 710.3  & -0.54  & -1000.00  & 50  & 0 \\ 
 090309B  &  60.0  & 4.70e+00  & 4.43e+00  & PL+HEC  & 197.0  & -1.52  & -1000.00  & 26  & 0 \\ 
  090310  & 125.2  & 2.15e+00  & 4.40e+00  & PL+HEC  & 279.0  & -0.65  & -1000.00  & 77  & 0 \\ 
  090319  &  67.7  & 7.47e+00  & 3.85e+00  & PL+HEC  & 187.3  &  0.90  & -1000.00  & 27  & 0 \\ 
 090320A  &  10.0  & -1.00e+00  & -1.00e+00  &      *  &  -1.0  & 1000.00  & -1000.00  & 60  & 0 \\ 
 090320B  &  52.0  & 1.10e+00  & 1.20e-01  & PL+HEC  &  72.0  & -1.10  & -1000.00  & 101  & 0 \\ 
 090320C  &   4.0  & -1.00e+00  & -1.00e+00  &      *  &  -1.0  & 1000.00  & -1000.00  & 40  & 0 \\ 
  090323  &  70.0  & 1.00e+02  & 1.23e+01  & PL+HEC  & 697.0  & -0.89  & -1000.00  & -1  & 1 \\ 
  \hline
 \end{tabular}
 \end{table}
 \pagebreak
 \newpage
\begin{table}[t]
\begin{tabular}{llllllllll}
 \hline
GRB name &   t90 &     fluence &       peakflux &    Function & E$_{\rm peak}$ &  $\alpha$ &    $\beta$ &       $\theta$ &  LAT\\
                     &    sec &      $10^{-6}$erg/cm$^2$& ph/cm$^2$/sec& keV & & & \\ 
\hline
 090326  &  11.2  & 8.60e-01  & -1.00e+00  & PL+HEC  &  75.0  & -0.86  & -1000.00  & 103  & 0 \\ 
  090327  &  24.0  & 3.00e+00  & 3.50e+00  &   Band  &  89.7  & -0.39  & -2.90  & 66  & 0 \\ 
 090328A  & 100.0  & 8.09e+01  & 1.85e+01  &   Band  & 653.0  & -0.93  & -2.20  & -1  & 1 \\ 
 090328B  &   0.3  & 9.61e-01  & 2.98e+01  &   Band  & 1967.0  & -0.92  & -2.48  & 74  & 0 \\ 
  090330  &  80.0  & 1.14e+01  & 6.80e+00  &   Band  & 246.0  & -0.99  & -2.68  & 50  & 0 \\ 
  090331  &  -1.0  & -1.00e+00  & -1.00e+00  &      *  &  -1.0  & 1000.00  & -1000.00  & 40  & 0 \\ 
  090403  &  16.0  & -1.00e+00  & -1.00e+00  &      *  &  -1.0  & 1000.00  & -1000.00  & 42  & 0 \\ 
  090405  &   1.2  & -1.00e+00  & -1.00e+00  &      *  &  -1.0  & 1000.00  & -1000.00  & 70  & 0 \\ 
  090409  &  20.0  & 6.14e-01  & 1.36e+00  & PL+HEC  & 137.0  &  1.20  & -1000.00  & 90  & 0 \\ 
 090411A  &  24.6  & 8.60e+00  & 3.25e+00  &   Band  & 141.0  & -0.88  & -1.82  & 59  & 0 \\ 
 090411B  &  18.7  & 8.00e+00  & 7.40e+00  &   Band  & 189.0  & -0.80  & -2.00  & 111  & 0 \\ 
  090412  &   0.5  & -1.00e+00  & -1.00e+00  &      *  &  -1.0  & 1000.00  & -1000.00  & 71  & 0 \\ 
 090418C  &   0.6  & 6.00e-01  & 8.50e+00  & PL+HEC  & 1000.0  & -0.94  & -1000.00  & 58  & 0 \\ 
  090422  &  10.0  & 1.00e+00  & 7.80e+00  &    SPL  &  -1.0  &  1.81  & -1000.00  & 29  & 0 \\ 
  090423  &  12.0  & 1.10e+00  & 3.30e+00  & PL+HEC  &  82.0  & -0.77  & -1000.00  & 75.6  & 0 \\ 
  090424  &  52.0  & 5.20e+01  & 1.37e+02  &   Band  & 177.0  &  0.90  & -2.90  & 71  & 0 \\ 
  090425  &  72.0  & 1.30e+01  & 1.40e+01  &   Band  &  69.0  & -1.29  & -2.03  & 105  & 0 \\ 
 090426B  &   3.8  & 5.20e-01  & -1.00e+00  &    SPL  &  -1.0  & -1.60  & -1000.00  & 56  & 0 \\ 
 090426C  &  12.0  & 3.10e+00  & 6.80e+00  &   Band  & 295.0  & -1.29  & -1.98  & 69  & 0 \\ 
 090427B  &   7.0  & 8.00e-01  & -1.00e+00  &    SPL  &  -1.0  & -1.10  & -1000.00  & 14  & 0 \\ 
 090427C  &  12.5  & 1.60e+00  & -1.00e+00  & PL+HEC  &  75.0  &  0.35  & -1000.00  & 81  & 0 \\ 
 090428A  &   8.0  & 9.90e-01  & 1.23e+01  &   Band  &  85.0  & -0.40  & -2.70  & 96  & 0 \\ 
 090428B  &  30.0  & 5.20e+00  & 1.01e+01  &   Band  &  53.0  & -1.81  & -2.17  & 101  & 0 \\ 
 090429C  &  13.0  & 3.70e+00  & 6.70e+00  &    SPL  &  -1.0  & -1.43  & -1000.00  & 112  & 0 \\ 
 090429D  &  11.0  & 1.60e+00  & 8.60e-01  &    SPL  & 223.0  & -0.87  & -1000.00  & 33  & 0 \\ 
  090502  &  66.2  & 3.50e-02  & 6.20e+00  & PL+HEC  &  63.2  & -1.10  & -1000.00  & 77  & 0 \\ 
  090509  & 295.0  & 8.40e+00  & 3.10e+00  & PL+HEC  & 343.0  & -0.90  & -1000.00  & 75  & 0 \\ 
 090510A  &   1.4  & 3.00e+01  & 8.00e+01  &   Band  & 4400.0  & -0.80  & -2.60  & -1  & 1 \\ 
 090510B  &   7.0  & -1.00e+00  & -1.00e+00  &      *  &  -1.0  & 1000.00  & -1000.00  & 100  & 0 \\ 
  090511  &  14.0  & 1.80e+00  & 2.50e+00  & PL+HEC  & 391.0  & -0.95  & -1000.00  & 67  & 0 \\ 
 090513A  &  23.0  & 6.80e+00  & 2.70e+00  & PL+HEC  & 850.0  & -0.90  & -1000.00  & 89  & 0 \\ 
 090513B  &  -1.0  & -1.00e+00  & -1.00e+00  &      *  &  -1.0  & 1000.00  & -1000.00  & 119  & 0 \\ 
  090514  &  49.0  & 8.10e+00  & 7.60e+00  &    SPL  &  -1.0  & -1.92  & -1000.00  & 19  & 0 \\ 
 090516A  & 350.0  & 2.30e+01  & 5.30e+00  &   Band  &  51.4  & -1.03  & -2.10  & 20  & 0 \\ 
 090516B  & 350.0  & 3.00e+01  & 4.00e+00  & PL+HEC  & 327.0  & -1.01  & -1000.00  & 45  & 0 \\ 
 090516C  &  15.0  & 4.00e+00  & 7.70e+00  &   Band  &  38.0  & -0.44  & -1.81  & 69  & 0 \\ 
 090518A  &   9.0  & 1.60e+00  & 4.70e+00  &    SPL  &  -1.0  & -1.59  & -1000.00  & 53  & 0 \\ 
 090518B  &  12.0  & 2.20e+00  & 5.60e+00  & PL+HEC  & 127.0  & -0.74  & -1000.00  & 90  & 0 \\ 
 090519B  &  87.0  & 1.40e+00  & 5.02e+00  &    SPL  &  -1.0  & -1.63  & -1000.00  & 18  & 0 \\ 
 090520B  &   1.5  & 4.50e-01  & 4.10e+00  &    SPL  &  -1.0  & -1.40  & -1000.00  & 10  & 0 \\ 
 090520C  &   4.9  & 3.54e+00  & 4.47e+00  &   Band  & 204.2  & -0.73  & -1.96  & 71  & 0 \\ 
 090520D  &  12.0  & 4.00e+00  & 4.10e+00  &   Band  &  46.3  & -0.99  & -3.25  & 66  & 0 \\ 
  090522  &  22.0  & 1.20e+00  & 3.50e+00  & PL+HEC  &  75.8  & -1.03  & -1000.00  & 53  & 0 \\ 
  090524  &  72.0  & 1.85e+01  & 1.41e+01  &   Band  &  82.6  & -1.00  & -2.30  & 63  & 0 \\ 
 090528A  &  68.0  & 9.30e+00  & 7.60e+00  & PL+HEC  &  99.0  & -1.70  & -1000.00  & 81  & 0 \\ 
 090528B  & 102.0  & 4.65e+01  & 1.47e+01  &   Band  & 172.0  & -1.10  & -2.30  & 65  & 0 \\ 
  \hline
 \end{tabular}
 \end{table}
 \pagebreak
 \newpage
\begin{table}[t]
\begin{tabular}{llllllllll}
 \hline
GRB name &   t90 &     fluence &       peakflux &    Function & E$_{\rm peak}$ &  $\alpha$ &    $\beta$ &       $\theta$ &  LAT\\
                     &    sec &      $10^{-6}$erg/cm$^2$& ph/cm$^2$/sec& keV & & & \\ 
\hline
  090529B  &   5.1  & 3.40e-01  & 4.10e+00  &   Band  & 142.0  & -0.70  & -2.00  & 36  & 0 \\ 
 090529C  &  10.4  & 3.10e+00  & 2.50e+01  &   Band  & 188.0  & -0.84  & -2.10  & 69  & 0 \\ 
 090530B  & 194.0  & 5.90e+01  & 1.08e+01  &   Band  &  67.0  & -0.71  & -2.42  & 84  & 0 \\ 
 090531B  &   2.0  & 6.20e-01  & 1.49e+00  &   Band  & 2166.0  & -0.71  & -2.47  & 25  & 0 \\ 
  090602  &  16.0  & 5.70e+00  & 3.62e+00  & PL+HEC  & 503.0  & -0.56  & -1000.00  & 112  & 0 \\ 
  090606  &  60.0  & 3.19e+00  & 2.41e+00  &    SPL  &  -1.0  & -1.63  & -1000.00  & 128  & 0 \\ 
  090608  &  61.0  & 3.20e+00  & 2.70e+00  &    SPL  &  -1.0  & -1.83  & -1000.00  & 93  & 0 \\ 
 090610A  &   6.5  & 7.32e-01  & 9.40e-01  &    SPL  &  -1.0  & -1.30  & -1000.00  & 70  & 0 \\ 
 090610B  & 202.5  & 4.13e+00  & 1.54e+00  & PL+HEC  & 104.9  & -0.46  & -1000.00  & 91  & 0 \\ 
 090610C  &  18.1  & 8.54e-01  & 1.12e+00  &    SPL  &  -1.0  & -1.62  & -1000.00  & 104  & 0 \\ 
  090612  &  58.0  & 2.37e+00  & 1.63e+00  &   Band  & 357.0  & -0.60  & -1.90  & 56  & 0 \\ 
  090616  &   2.7  & 2.23e-01  & 2.08e+00  &    SPL  &  -1.0  & -1.27  & -1000.00  & 68  & 0 \\ 
  090617  &   0.4  & 4.68e-01  & 1.00e+01  &   Band  & 684.0  & -0.45  & -2.00  & 45  & 0 \\ 
  090618  & 155.0  & 2.70e+02  & 7.34e+01  &   Band  & 155.5  & -1.26  & -2.50  & 133  & 0 \\ 
  090620  &  16.5  & 6.60e+00  & 7.00e+00  &   Band  & 156.0  & -0.40  & -2.44  & 60  & 0 \\ 
 090621A  & 294.0  & 4.40e+00  & 1.92e+00  &   Band  &  56.0  & -1.10  & -2.12  & 12  & 0 \\ 
 090621B  &   0.1  & 3.71e-01  & 6.40e+00  &   Band  & 321.6  & -0.13  & -1.57  & 108  & 0 \\ 
 090621C  &  59.9  & 1.80e+00  & 2.29e+00  & PL+HEC  & 148.0  & -1.40  & -1000.00  & 52  & 0 \\ 
 090621D  &  39.9  & 1.34e+00  & 1.74e+00  &    SPL  &  -1.0  & -1.66  & -1000.00  & 79  & 0 \\ 
  090623  &  72.2  & 9.60e+00  & 3.30e+00  &   Band  & 428.0  & -0.69  & -2.30  & 73  & 0 \\ 
 090625A  &  51.0  & 8.80e-01  & 5.00e-01  & PL+HEC  & 198.0  & -0.60  & -1000.00  & 13  & 0 \\ 
 090625B  &  13.6  & 1.04e+00  & 1.87e+00  &   Band  & 100.0  & -0.40  & -2.00  & 125  & 0 \\ 
  090626  &  70.0  & 3.50e+01  & 1.79e+01  &   Band  & 175.0  & -1.29  & -1.98  & 15  & 1 \\ 
  090630  &   5.1  & 5.10e-01  & 2.78e+00  &   Band  &  71.0  & -1.50  & -2.30  & 75  & 0 \\ 
  090701  &  12.0  & 4.50e-01  & 2.10e+00  &    SPL  &  -1.0  &  1.84  & -1000.00  & 13  & 0 \\ 
  090703  &   9.0  & 6.80e-01  & 1.00e+00  &    SPL  &  -1.0  & -1.72  & -1000.00  & 25  & 0 \\ 
  090704  &  70.0  & 5.80e+00  & 1.20e+00  & PL+HEC  & 233.7  & -1.13  & -1000.00  & 77  & 0 \\ 
  090706  & 100.0  & 1.50e+00  & 1.24e+00  &    SPL  &  -1.0  & -2.16  & -1000.00  & 20  & 0 \\ 
  090708  &  18.0  & 4.00e-01  & 1.00e+00  & PL+HEC  &  47.5  & -1.29  & -1000.00  & 55  & 0 \\ 
 090709B  &  32.0  & 1.30e+00  & 2.00e+00  & PL+HEC  & 130.0  & -1.01  & -1000.00  & 35  & 0 \\ 
  090711  & 100.0  & 1.17e+01  & 4.20e+00  & PL+HEC  & 210.0  & -1.30  & -1000.00  & 13  & 0 \\ 
  090712  &  72.0  & 4.20e+00  & 6.30e-01  & PL+HEC  & 505.0  & -0.68  & -1000.00  & 33  & 0 \\ 
  090713  & 113.0  & 3.70e+00  & 1.60e+00  & PL+HEC  &  99.0  & -0.34  & -1000.00  & 63  & 0 \\ 
 090717A  &  70.0  & 4.50e-01  & 7.80e+00  &   Band  & 120.0  & -0.88  & -2.33  & 70  & 0 \\ 
 090717B  &   0.9  & 4.83e-01  & 3.91e+00  &    SPL  &  -1.0  & -1.02  & -1000.00  & 35  & 0 \\ 
 090718A  &  -1.0  & -1.00e+00  & -1.00e+00  &      *  &  -1.0  & 1000.00  & -1000.00  & 51  & 0 \\ 
 090718B  &  28.0  & 2.52e+01  & 3.20e+01  &   Band  & 184.0  & -1.18  & -2.59  & 76  & 0 \\ 
 \hline
 \end{tabular}
 \end{table}
 \pagebreak
 \newpage
\begin{table}[t]
\begin{tabular}{llllllllll}
 \hline
GRB name &   t90 &     fluence &       peakflux &    Function & E$_{\rm peak}$ &  $\alpha$ &    $\beta$ &       $\theta$ &  LAT\\
                     &    sec &      $10^{-6}$erg/cm$^2$& ph/cm$^2$/sec& keV & & & \\ 
\hline
  090719  &  16.0  & 4.83e+01  & 3.78e+01  &   Band  & 254.0  & -0.68  & -2.92  & 88  & 0 \\ 
 090720A  &   7.0  & 2.90e+00  & 1.09e+01  & PL+HEC  & 117.5  & -0.75  & -1000.00  & 113  & 0 \\ 
 090720B  &  20.0  & 1.06e+01  & 1.09e+01  &   Band  & 924.0  & -1.00  & -2.43  & 56  & 0 \\ 
 090802A  &   0.1  & 6.50e-01  & 6.12e+01  &   Band  & 283.0  & -0.42  & -2.40  & 123  & 0 \\ 
 090802B  &  -1.0  & -1.00e+00  & -1.00e+00  &      *  &  -1.0  & 1000.00  & -1000.00  & 104  & 0 \\ 
 090807B  &   3.0  & 1.02e+00  & 1.09e+01  &   Band  &  37.0  & -0.60  & -2.40  & 45  & 0 \\ 
 090809B  &  15.0  & 2.26e+01  & 2.36e+01  &   Band  & 198.0  & -0.85  & -2.02  & 81  & 0 \\ 
  090813  &   9.0  & 3.50e+00  & 1.44e+01  &   Band  &  95.0  & -1.25  & -2.00  & 35.3  & 0 \\ 
 090814C  &   0.2  & 6.60e-01  & 9.10e+00  & PL+HEC  & 790.0  & -0.39  & -1000.00  & 61  & 0 \\ 
 090815A  & 200.0  & 3.40e+00  & 1.90e+00  &    SPL  &  -1.0  & -1.50  & -1000.00  & 87  & 0 \\ 
 090815B  &  30.0  & 5.05e+00  & 1.44e+01  &   Band  &  18.6  & -1.82  & -2.70  & 82  & 0 \\ 
  090817  & 220.0  & 7.30e+00  & 3.80e+00  &   Band  & 115.0  & -1.10  & -2.20  & 82  & 0 \\ 
 090820A  &  60.0  & 6.60e+01  & 5.80e+01  &   Band  & 215.0  & -0.69  & -2.61  & 108  & 0 \\ 
 090820B  &  11.2  & 1.16e+00  & 6.10e+00  & PL+HEC  &  38.8  & -1.44  & -1000.00  & 32  & 0 \\ 
  090826  &   8.5  & 1.26e+00  & 3.28e+00  & PL+HEC  & 172.0  & -0.96  & -1000.00  & 35  & 0 \\ 
  090828  & 100.0  & 2.52e+01  & 1.62e+01  &   Band  & 136.5  & -1.23  & -2.12  & 95  & 0 \\ 
 090829A  &  85.0  & 1.02e+02  & 5.15e+01  &   Band  & 183.0  & -1.44  & -2.10  & 47  & 0 \\ 
 090829B  & 100.0  & 6.40e+00  & 3.20e+00  &   Band  & 143.0  & -0.70  & -2.40  & 42  & 0 \\ 
  090831  &  69.1  & 1.66e+01  & 9.40e+00  &   Band  & 243.8  & -1.52  & -1.96  & 107  & 0 \\ 
 090902A  &   1.2  & 2.11e+00  & 1.14e+01  &   Band  & 388.0  &  0.30  & -2.05  & 82  & 0 \\ 
 090902B  &  21.0  & 3.74e+02  & 4.61e+01  &   Band  & 798.0  & -0.61  & -3.87  & 52  & 1 \\ 
 090904B  &  71.0  & 2.44e+01  & 9.80e+00  &   Band  & 106.3  & -1.26  & -2.18  & 113  & 0 \\ 
  090910  &  62.0  & 9.20e+00  & 2.30e+00  &   Band  & 274.8  & -0.90  & -2.00  & 107  & 0 \\ 
  090922A  &  92.0  & 1.14e+01  & 1.56e+01  &   Band  & 139.3  & -0.77  & -2.28  & 19  & 0 \\ 
  090925  &  50.0  & 9.46e+00  & 4.20e+00  &   Band  & 156.0  & -0.60  & -1.91  & 116  & 0 \\ 
 090926A  &  20.0  & 1.45e+02  & 8.08e+01  &   Band  & 268.0  & -0.69  & -2.34  & 52  & 1 \\ 
 090926B  &  81.0  & 8.70e+00  & -1.00e+00  & PL+HEC  &  91.0  & -0.13  & -1000.00  & 100  & 0 \\ 
  090927  &   2.0  & 6.10e-01  & 7.20e+00  &    SPL  &  -1.0  & -1.47  & -1000.00  & 85  & 0 \\ 
 090929A  &   8.5  & 1.06e+01  & 1.09e+01  & PL+HEC  & 610.9  & -0.52  & -1000.00  & 122  & 0 \\ 
 091003A  &  21.1  & 3.76e+01  & 3.18e+01  &   Band  & 486.2  & -1.13  & -2.64  & 13  & 1 \\ 
\label{t:fit}
\end{tabular}
\end{table}

\begin{table}[t]
\caption{ 
Best fit parameters $Rate(z=0)$ , $L^*$, $\alpha$ and
$\beta$ and their 1-$\sigma$ confidence levels. For each fit we
report the  $\chi^2$ values corresponding to the best fit
($\chi^2_{\rm b.f.}$). Also shown are the parameter for a LF,
LFb, that fit quite well all the samples, the $\chi^2\sim 1.3 $ for all the instruments }
\begin{tabular}{|c|c|c|c|c|c|}
 \hline
sample
& Rate(z=0)  & $L^*$ & $\alpha$ & $\beta$ &$\chi^2$\\
   & $Gpc^{-3} yr^{-1}$ & $10^{51}$ erg/sec &  & &   \\
  \hline
GBM & $0.5^{+0.3}_{-0.2}$ & $5.5^{+1.5}_{-2}$ & $0.3^{+0.1}_{-0.5}$
&$2.3^{+0.6}_{-0.3}$ & 1.1  \\
BATSE & $1.0^{+0.2}_{-0.4}$ & $4_{-1.5}^{+2}$ &
$0.1_{-0.05}^{+0.3}$ &$2.6_{-0.5}^{+0.9}$ & 1.1 \\
{\it Swift} & $0.6^{+0.3}_{-0.1}$ & $3.3_{-0.5}^{+2.5}$ &
$0.1_{-0.05}^{+0.3}$ &$2.7_{-0.4}^{+1}$ & 0.95 \\
LFb &   $0.8$  & $5 $ &
$0.1$ & $2.7$ & 1.3 \\
  \hline
\end{tabular}
\end{table}

\section{Acknowledgements}

We are grateful to Eli Waxman for very helpful comments and discussions. We  thank Nicola Omodei, Toby Burnett and David Croward for very useful discussions.

\end{document}